\documentclass[pra,preprint,superscriptaddress,tightenlines,12pt]{revtex4-1} 
\usepackage{epsfig} 
\usepackage{amsmath,amssymb}

\usepackage{graphicx}
\usepackage{dcolumn}
\usepackage{bm}
\usepackage{hyperref}
\usepackage{bookmark}
\usepackage{color}

\begin{document} 
  
\title{\texorpdfstring{$\beta$}{Lg}-electron spectrum: static screened Coulomb field and exchange effects } 
                                      
\author{B. Najjari$^\text{1}$}
\email{bennaceur.najjari@iphc.cnrs.fr}
\author{X. Mougeot$^\text{2}$}
\email{xavier.mougeot@cea.fr}
\author{M.-M. B\'e$^\text{2}$}
\author{C. Bisch$^\text{2}$}
\author{P.-A. Hervieux$^\text{3}$}
\author{A. Nachab$^\text{4}$}
\author{A.-M. Nourreddine$^\text{1}$}

\affiliation{ $^\text{1}$Institut Pluridisciplinaire Hubert Curien, CNRS et
Universit\'e de Strasbourg,
23 rue du Loess, BP 28, 67037 Strasbourg Cedex 2, France } 
\affiliation{$^\text{2}$CEA, LIST, Laboratoire National Henri Becquerel (LNE-LNHB), F-91191 Gif-sur-Yvette, France}
\affiliation{$^\text{3}$Institut de Physique et Chimie des Mat\'eriaux de Strasbourg,
 CNRS et Universit\'e de Strasbourg, BP 43, F-67034 Strasbourg Cedex, France}
\affiliation{$^\text{4}$ D\'epartement de physique, Facult\'e Poly-disciplinaire de Safi, Universit\'e Cadi Ayyad, Route Sidi Bouzid BP 4162, 46000 Safi, Maroc}
 
\date{\today}  

\begin{abstract} 
We consider the energy spectrum of emitted electrons in $\beta$-decay. Exact Coulomb Dirac wave functions describing the $\beta$-electron in the Coulomb field of the daughter nucleus are used. Further, the improved wave functions which include the screening of the Coulomb field due to the atomic electron cloud are also used. Thus, the interaction between the $\beta$-electron and the field due to the daughter atom is treated within a nonperturbative approach. Are shown the modifications due to the screening on the $\beta$ spectra and shown that those effects are very important. In addition, are addressed the contributions to the $\beta$ spectra due to the exchange terms and shown that the corresponding effects can be substantial. Higher orders arising from the multipole expansion are considered. A comparison of the theoretical results obtained in this work has been made with recent experimental data and a very good agreement was observed.
\end{abstract}


\maketitle  

\section{Introduction}

A $\beta$-decay in a radionuclide represents an interesting and fascinating process where an excess energy is converted into matter and vice versa. For instance, in the process of $\beta$-electron emission, a neutron is converted into a proton by getting rid of the energy excess via the emission of an electron and an antineutrino. On the other hand, $\beta$-decay has been attracted the attention of different physical communities. For example, it has been considered as a powerful tool to study nuclear spectroscopy and obtained fundamental importance in nuclear and elementary particle physics and played an important key role for our understanding of fundamental interactions and their properties related to the symmetry and the conservation law (see \cite{herwig-schopper}, \cite{behrens}). Besides, a precise knowledge of the shapes of the $\beta$ spectra would substantially reduce the uncertainties in the measurements of the activity in ionizing radiation metrology. Furthermore, these spectra are of strong interest in nuclear power plants and medical therapy. For example, in medical therapy an accurate spectrum of emitted electrons in a $\beta$ process is crucial in the determination of necessary doses which could be delivered to a given patient.

Owing to the recent experimental advances as well as the increasing accuracy achieved in $\beta$-decay measurements, more theoretical studies in $\beta$-decay processes have been strongly stimulated (see \textit{e.g.} \cite{mougeot} and references therein). Despite some theoretical activities during the last decades the field of the $\beta$-decay remains largely unexplored. These studies which have been carried out were mainly focused on allowed and first forbidden nonunique transitions. Thus, it is particularly appropriate to address a systematic and a complete theoretical investigation extended to higher orders in the multipole expansion including the forbidden unique and nonunique transitions.

A detailed nonperturbative treatment of the static Coulomb field arising from the appropriate daughter nucleus charge have been presented by Halpern \cite{halpern}. In this study, the effects related to the static Coulomb field of the daughter nucleus have been included in the wave functions of the emitted $\beta$-electron. In addition, the shape of the $\beta$ spectra, mainly at low energies, can be affected by the presence of the atomic electron cloud. Thus, a nucleus experiencing a $\beta$-decay is generally surrounded by an atomic electron cloud and, in particular, a number of corrections due to the atomic electrons might influence the spectrum of emitted $\beta$-particle. These corrections arise through the inclusion of the screening in the wave function of the emitted $\beta$-particle or through exchange processes involving bound and continuum electrons \cite{pype-and-harston} if $\beta$-electron is involved. The corresponding effects are more important at relatively soft energy emissions. The first study of the exchange effect was limited to the $1s$ orbital. It has been shown that the exchange term in the transition matrix element could interfere with the direct term and therefore can lead to a lower probability emission at low energy \cite{bahcall}.

The main aim of the present contribution is to address the above {\it atomic} effects, namely screening and exchange whatever the transition order. First we start with the definition of the $\beta$-decay form factors related to the nuclear and lepton vector currents. For such a consideration of nuclear $\beta$-decay, we will use the so-called \textit{phenomenological} form of the weak interaction theory \cite{behrens}. Hence, the effects related to the structure of the nucleons or meson exchange are ignored. We will mainly focus our attention on the leptonic current involving the $\beta$-electron and antineutrino wave functions. While, the nuclear current will be considered in the simplest approach, where the nucleons are described within the framework of Dirac plane waves. Although this approximation is crude, it enables one to address the relevant features and a qualitative understanding of the implications of the multipole expansion in the transition matrix element \cite{behrens}. Effects due to the static Coulomb field of the nucleus as well as the screening corrections associated to the atom electron cloud are included in the nonperturbative treatment. Therefore the \textit{post} electromagnetic interaction between the $\beta$-electron and the daughter nucleus is taken into account to all orders.

This paper is organized as follows. In Sec. \ref{secform}, we present the theory for $\beta$-decay based on the simple weak phenomenological interaction and establish the conventions and definitions used in this work. The description of the $\beta$-electron wave function in a screened potential due to the atomic electron cloud is discussed. The antisymmetrization in the final-state wave function under the exchange of the $\beta$-electron and the atomic electron coordinates, resulting in an exchange contribution in the transition amplitude is addressed in this section. In Sec. \ref{secres}, we compare the results of our calculations with some experimental data and discuss some aspects of the energy spectra of emitted $\beta$-electron. Atomic units are used throughout except where otherwise stated.

\section{General considerations}
\label{secform}

In $\beta$-decay, a radionuclide undergoes a change in nuclear charge by one unit. This {\it sudden} change in the nuclear charge could lead to a modification in the atomic electron cloud, as a result an atomic electron may undergo an excitation or ionization transitions, due to the imperfect overlap in the atomic parent and atomic daughter wave functions. These modifications are negligible ($ \le 0.01 \% $) \cite{mougeot} and hence can be ignored in this work.
However, in the present work we consider some atomic electron effects, namely: 

\noindent $i)$ the screening effect, where the atomic electron cloud can partially screen the Coulomb field of the daughter nucleus, 

\noindent $ii)$ the exchange effect which is a consequence of the antisymmetrization in the final-state wave function under the exchange of the $\beta$-electron and the atomic electron coordinates. This exchange leads to the creation of a $\beta$-electron into a bound orbital of the daughter atom resulting in a simultaneous emission of a bound atomic electron into a continuum state of the same daughter atom.

\subsection{The scattering matrix}

To describe a $\beta$-decay process of a radionuclide which is subject to weak interaction, one can start with the consideration of the S-matrix given in the point-like nature of the decay, according to Fermi's theory, by (see \textit{e.g.} \cite{mandl-and-shaw}),

\begin{eqnarray}
\hat S = - \dfrac{4\pi i}{c^2} \dfrac{G_\beta}{\sqrt{2}}
\int d^4\kappa  {\cal L}_\mu(\kappa) {\cal D}^\mu(-\kappa).
\label{1}
\end{eqnarray}
In this expression, and similarly everywhere after, the summation over repeated greek letters is assumed. The covariant $ a_\mu$ and contravariant $a^\mu$ four-vectors are connected by $a_\mu = g_{\mu \nu} a^\nu $, where $g_{\mu \nu}$ is the metric tensor of the four-dimensional flat-space and defined by $g_{00}=-g_{11}=-g_{22}=-g_{33} = 1$ and $g_{\mu \nu}=0$  for ${\mu \neq \nu}$. $G_\beta = G \cos \Theta$ stands for the vector coupling constant in the weak interaction, with $G$ the universal weak coupling constant and $\Theta$ the Cabbibo angle. $c$ denotes the speed of light. Further, the four-vectors $ {\cal L}_\mu $ and $ {\cal D}_\mu$ ($\mu = 0, 1 \dots 3 $) are the lepton and nuclear form factors, respectively, and are given in the context of a pure V-A weak interaction by: 
\begin{eqnarray}
{\cal L}^\mu (\kappa) = \dfrac{1}{4\pi^2}
\int d^4x \  \text{e}^{-i \kappa x } \left[ \bar{\psi_e}(x) \gamma^\mu (1-\gamma_5)\psi_{\nu_e}(x)\right]
\label{1a}
\end{eqnarray}
and 
\begin{eqnarray}
{\cal D}^\mu (-\kappa)= \dfrac{1}{4\pi^2}
\int d^4y \  \text{e}^{i \kappa y } \left[ \bar{\psi_p}(y) \gamma^\mu (1+g\gamma_5) \psi_{n}(y)\right]
\label{1b}
\end{eqnarray}
where $g=1.25$ is the ratio $G_A / G_V$ where $G_A$ and $G_V$ are respectively the axial vector and vector coupling constants in the weak interaction. In the case of point-like nucleons, \textit{i.e.} without any internal structure, $g=1$, but, throughout this work we have used $g=1.25$. The quantities $\bar{\psi_e} $, $ \psi_{\nu_e} $, $ \bar{\psi_p} $ and $ \psi_{n} $ stand for the creation and destruction field operators of the corresponding particle. The $\gamma^\mu$ with ($\mu = 0, 1\dots 3$) and $\gamma_5$ \cite{gammas} are the standard Dirac matrices (see \cite{mandl-and-shaw}).  

The transition S-matrix element taken between the initial $\left| i \right>$ and final $\left| f \right>$ nuclear states of the nucleon reads $S_{fi} = \left<f\left|\hat S \right| i\right>$. In the case of $\beta$-electron emission  the initial and final states (in the Fock-space) during the process are given, respectively, by  $\left| i \right>  = \left|n,0,0,0 \right> $ and $\left| f \right>  = \left|0,p,e^-,\bar\nu_e \right> $.

The wave functions in real space are obtained by acting the creation and destruction operator on the Fock-states as follow:
\[
\bar{\psi_e}(x) \left|0\right>= 
\left|e\right>\bar{\varphi_e}({\bf r}) {\text e}^{i E_e t},
\]
\[
\psi_{\nu_e}(x) \left|0\right>= 
\left|{\bar\nu_e}\right>\varphi_{\bar\nu_e}({\bf r}) {\text e}^{i E_{\bar\nu_e} t},
\]
\[
\bar\psi_{p}(y) \left|0\right>= 
\left|p\right>\bar\chi_p({\bf r'}) {\text e}^{i E_p t'}
\]
and
\[
\psi_{n}(y) \left|n\right>= 
\left|0\right>\chi_n({\bf r'}) {\text e}^{-i E_n t'}.
\]
$\varphi_e$ describes the wave function, in real space, of the emitted $\beta$-electron with energy $E_e$ moving in the static Coulomb field of the daughter nucleus and in presence of the atomic electron cloud. While $\varphi_{\bar\nu_e}$ stands for the antineutrino wave function with a positive energy $E_{\bar\nu_e}$. Similarly, the proton and neutron are described by $\chi_p$ and $\chi_n$ with energies $E_p$ and $E_n$, respectively. Inserting these expressions in equation (\ref{1}) and after performing the integration over time component $\kappa_0$ of $\kappa = (\kappa_0, \bf k )$ one can easily obtain 
\begin{widetext}
\begin{eqnarray}
{S}_{fi}\!&=&\!- \dfrac{i}{\pi c} \dfrac{G_\beta}{\sqrt{2}}
 \delta(E_n\!-\!E_p\!-\!E_e-E_{\bar\nu_e}) 
 \int d{\bf k} L^\mu({\bf k})
D_\mu(-{\bf k}).
\label{S-matrix-element}
\end{eqnarray}
\end{widetext}
In the above equation the $\delta$ function ensures the energy conservation. The four-vector form factors $L^\mu$ and $D^\mu$ are given by:
\begin{eqnarray}
L^\mu({\bf k}) = 
\int d{\bf r} \  \text{e}^{i {\bf k \cdot r} } \bar{\varphi_e}({\bf r}) 
\gamma^\mu (1-\gamma_5) \varphi_{\bar\nu_e}({\bf r})
\label{L-vect}
\end{eqnarray}
and
\begin{eqnarray}
D^\mu(-{\bf k})=
\int d{\bf r'} \  \text{e}^{-i{\bf k \cdot r'} }  \bar{\chi_p}({\bf r'}) 
\gamma^\mu (1+g\gamma_5) \chi_{n}({\bf r'}).
\label{D-vect}
\end{eqnarray}

Before evaluating these integrals it is worth to stress the following points: 

\noindent $i)$ Due to the large mass of the nucleus, the typical recoil velocities of the nucleus after the decay in the laboratory reference frame are not only nonrelativistic but much below the $\beta$-electron velocity. Taking this into account, the laboratory frame and the center of the nucleus can be assumed as identical. 

\noindent $ii)$ For simplicity, the finite size of the nucleus is ignored and is assumed as point-like with charge $Z$.

\subsection{Simplification of the nuclear current}

The motion of the nucleons, during the decay, is assumed to be a free motion and thus is described by free Dirac four-spinors. A better description of the motion of the nucleons inside the nucleus is beyond the scope of the present paper and will be given in a future paper. All other effects related to the structure of the nucleons and meson exchange are also ignored here. 
Accordingly with their motion, the nucleons are assumed to be on their mass shell such that the proton and neutron wave functions are given by free Dirac plane waves: 

\[
 \chi_p({\bf r'}) = 
 \sqrt{\dfrac{m_pc^2}{V_p E_p}}
{\text e}^{i{\bf p.r'}}
 u({\bf p},s_p) 
\] 
\[
 \chi_n({\bf r'}) = 
 \sqrt{\dfrac{m_nc^2}{V_n E_n}}
{\text e}^{i{\bf n.r'}}
 u({\bf n},s_n)
\]
where $V_p$ and $V_n$ stand for the normalization volumes for the Dirac plane waves describing the proton and neutron, respectively. $m_p$ and $m_n$ are the proton and neutron masses. $u({\bf k},s)$ is the Dirac free four-spinor of a given particle ("$p$" for a proton and "$n$" for a neutron) with momentum $\bf k$ and spin projection $s$. Inserting these expressions in the form factor $D^\mu(-{\bf k})$ one obtains easily 
\[
D^\mu(-{\bf k}) = 8\pi^3
\left(\dfrac{m_n m_p c^4}{V_p V_n E_p E_n } \right)^\frac{1}{2} 
 \delta({\bf n} - {\bf p} - {\bf k}) {\cal Q}_{{\bf p},{\bf n}}^\mu(s_p,s_n)
\]
with,
\[
{\cal Q}_{{\bf p},{\bf n}}^\mu(s_p,s_n) =
\bar{u}({\bf p},s_p) 
\gamma^\mu (1+g\gamma_5) u({\bf n},s_n).
\]
Inserting these results in the expression (\ref{S-matrix-element}) of the S-matrix element, accordingly the latter reads 
\begin{widetext}
\begin{eqnarray}
{S}_{fi}\!&=&\!- \dfrac{i8\pi^2}{c} \dfrac{G_\beta}{\sqrt{2}}
\left(\dfrac{m_n m_p c^4}{V_p V_n E_p E_n } \right)^\frac{1}{2}  
\delta(E_n\!-\!E_p\!-\!E_e\!-\!E_{\bar\nu_e}) 
L_\mu({\bf n} - {\bf p} )
{\cal Q}^\mu_{{\bf p},{\bf n}}(s_p,s_n).
\label{S-matrix-element2}
\end{eqnarray}
\end{widetext}
\subsubsection{Transition probability and decay rate}

Once the S-matrix element is obtained (\ref{S-matrix-element2}), the transition probability per unit of time ($W_{fi}$) can be derived following the standard way. Imposing ${\bf n} - {\bf p} = {\bf q} $ the transition probability reads: 

\begin{eqnarray}
W_{fi}  &=& \lim\limits_{T \to +\infty}\dfrac{\left|S_{fi}\right|^2}{T}\nonumber \\
        &=& \dfrac{G_\beta^2}{2}
\frac {32\pi^3 m_n m_p c^2}{V_p V_n E_p E_n } 
\delta(E_n-E_p-E_e-E_{\bar\nu_e}) 
 \left| L_\mu({\bf q} )
{\cal Q}^\mu_{\bf q} (s_p,s_n)\right|^2.
\end{eqnarray}
In the above expression we have used the relation $(\delta(E_n-E_p-E_e-E_{\bar\nu_e}) )^2 = \dfrac{T}{2\pi}\delta(E_n-E_p-E_e-E_{\bar\nu_e})$, with $T\rightarrow \infty $. The differential decay rate as a function of emitted $\beta$  electron is given by the relation: 
\[
\dfrac{d \Gamma}{d E_e} = \int 
\dfrac{ V_p d{\bf p}}{8\pi^3}
\dfrac{ V_\nu d{\bf p}_{\nu}}{8\pi^3}
 W_{fi}.
\] 
\subsection{Lepton form factor} 
In order to calculate the lepton form factor it is necessary to introduce the appropriate wave functions describing each lepton, namely neutrinos/antineutrinos and electrons/positrons. The neutrino (antineutrino) has no charge and interacts practically only through the weak interaction which is considered here as a perturbation causing the decay. Therefore, the unperturbed wave function (free of the weak interaction) describing the antineutrino is simply a Dirac plane wave and is readily given by:
\begin{eqnarray}
\varphi_{\bar\nu_e}({\bf r})=\sqrt{\dfrac{m_{\bar\nu_e}c^2}{VE_{\bar\nu_e}}} 
\text{e}^{-i {\bf p_{\bar\nu_e} \cdot r} } v({\bf p_{\bar\nu_e}},s_{\bar\nu_e})
\end{eqnarray}
$V$ stands for the normalization volume of the antineutrino wave function. ${\bf p_{\bar\nu_e}}$  and $s_{\bar\nu_e}$ are its momentum and spin projection, respectively, and $m_{\bar\nu_e}$ is its rest mass. $v({\bf p_{\bar\nu_e}},s_{\bar\nu_e})$ is the negative energy component of a neutrino with a momentum $-{\bf p_{\bar\nu_e}}$ and negative energy $-E_{\bar\nu_e}$.

The $\beta$-electron has an electric charge, thus its wave function may be distorted by the electromagnetic interaction with the nuclear charge and the atomic electron cloud. Therefore, the distorted wave function describing the $\beta$-electron with an angular momentum parameter $\kappa_e$ and a given energy $E_e$ can be written in the form \cite{rose}:

\begin{eqnarray}
 \varphi_e({\bf r}) = 
 \left(
\begin{array}{cc}
 g_{\kappa_e}(r) \,\chi_{\kappa_e}^{\mu_e}(\hat{\bf r})  \\
 \\
 i f_{\kappa_e}(r) \,\chi_{-\kappa_e}^{\mu_e}(\hat{\bf r}) \\
\end{array}
\right)
\label{electron-wf}
\end{eqnarray}
where
\[
\chi_\kappa^\mu(\hat{\bf r}) = \sum_{m_s=-1/2}^{+1/2} C^{j \mu}_{l\mu-m_s;\frac{1}{2}m_s} 
Y_{l\mu-m_s}(\hat{\bf r})\chi^{m_s}
\]
with $\chi^{1/2} = \left(^1_0 \right)$ and $\chi^{-1/2} = \left(^0_1 \right)$, the total kinetic momentum  $j = |\kappa|-1/2$ and the orbital momentum $l = |\kappa+1/2|-1/2$. $C^{j \mu}_{l\mu-m_s;\frac{1}{2}m_s}$ refer to the Clebsch-Gordan coefficients \cite{Varshalovitch}. Further, $g_{\kappa_e}$ and $f_{\kappa_e}$ are the large and small radial components of the electronic wave functions. For a given energy $E$, $g_{\kappa}$ and $f_{\kappa}$ are the solutions of the Dirac radial equations
\begin{eqnarray}
\left .
\begin{array}{r c l}
c\left( \dfrac{d}{dr} + \dfrac{\kappa +1}{r}\right)g_\kappa\!&=&\!\![E - V(r) + m_e c^2 ]f_{\kappa} \\ 
\\
c\left( \dfrac{d}{dr} - \dfrac{\kappa -1}{r}\right)f_\kappa\!&=&\!\!-[E - V(r) - m_e c^2  ]g_{\kappa}\\
\end{array}
\right \}
\label{Dirac-EQ}
\end{eqnarray}
where $m_e$ is the electron rest mass. The potential $V(r)$ is an electrostatic central atomic potential due to the {\it point-like} charge of the daughter nucleus embedded in the atomic electron cloud. This potential is described later in the next subsection.\\
The solutions of the Dirac radial equations (\ref{Dirac-EQ}) can be obtained by using standard numerical methods. Here we have followed the numerical method described in detail in \cite{salvat-radial} which is close to the method given in \cite{behrens}. Once the Dirac radial wave functions (\ref{electron-wf}) are obtained, the evaluation of the integral $ L^\mu({\bf k})$ (\ref{L-vect}) can be achieved. Hence it is very appropriate to proceed first in expanding, in equation (\ref{L-vect}), the plane waves in partial waves using the following relation:
\begin{eqnarray}
 \text{e}^{i {\bf q \cdot r} } = 4\pi \sum_{LM} i^L j_L(qr) Y_{LM}(\hat{\bf q}) Y^*_{LM}(\hat{\bf r})
 \nonumber
\end{eqnarray}
$L$ corresponds to the angular momentum carried by the radiation field and transmitted between the decaying nucleus and the arising leptons from the reaction. 

Each term in the above expansion refers to as a multipole of order $L$, and the lowest orders in this expansion are the most important ones due to the smallness of the product $qr$. The transitions arising from the first term in the expansion $L=0$ are referred to as {\it allowed} transitions and those associated to higher orders $L\geq 1$ are called {\it forbidden} ones. 
More precisely, the allowed transitions occur between two nuclear states with spin difference $\Delta J = |J_f-J_i| = 0, \: 1$ 
and having the same parity, i.e. $ \pi_i .  \pi_f = 1$, where $J_{i(f)}$ and $\pi_{i(f)}$ are the initial (final) spin and parity of the nucleus, respectively. However, forbidden transitions of order $L$ are occurring between states with $ \pi_i .  \pi_f = (-1)^L$ and spin difference 
$\Delta J =  L $ or $L+1$. The transitions with $\Delta J =  L $ are called nonunique  and those with $\Delta J =  L+1 $ are specified as being unique ones. Furthermore, since the nuclear states before and after the decay have well defined spins, the angular momentum transmitted to the lepton current is restricted to values which can be combined to the nuclear spins in such a way that angular momentum is conserved. Accordingly, we obtain for leptonic vector form factor 
\begin{eqnarray}
L^\mu({\bf k}) &=&\sqrt{\dfrac{m_{\bar\nu_e}c^2}{VE_{\bar\nu_e}}} (4\pi)^2 \sum_{LM}  
\sum_{l_\nu m_\nu} i^{L-l_\nu} Y_{LM}(\hat{\bf k}) 
Y^*_{l_\nu m_\nu}({\hat{\bf p}}_\nu)
{\cal M}^{LM}_{l_\nu m_\nu} (k,p_{\nu_e}) 
Q^\mu ({\bf p_{\bar\nu_e}},s_{\bar\nu_e}) \nonumber 
\end{eqnarray}
with, 
\begin{eqnarray}
Q^\mu ({\bf p_{\bar\nu_e}},s_{\bar\nu_e}) = 
\gamma^0\gamma^\mu (1-\gamma_5) v({\bf p_{\bar\nu_e}},s_{\bar\nu_e}) \nonumber
\end{eqnarray}
and 
\begin{eqnarray} 
{\cal M}^{LM}_{l_\nu m_\nu}(k,p_{\nu_e}) &=& \int d{\bf r} \  j_L(kr)   
j_{l_{\nu}}(p_{\nu_e} r )  \varphi_e^\dag({\bf r})
Y^*_{LM}(\hat{\bf r}) Y_{l_\nu m_\nu}({\hat{\bf r}}). \nonumber
\end{eqnarray}
Taking into account the expression (\ref{electron-wf}) of the above wave function of the emitted electron, ${\cal M}$ becomes 

\begin{eqnarray} 
{\cal M}^{LM}_{l_\nu m_\nu}(k,p_{\nu_e})  &=& \int d{\bf r}\ j_L(kr)    
j_{l_{\nu}}(p_{\nu_e} r ) 
 \left\{
\begin{matrix}
 g_{\kappa_e} \chi_{\kappa_e}^{\mu_e}(\hat{\bf r})  \\
 \\
 i f_{\kappa_e}\chi_{-\kappa_e}^{\mu_e}(\hat{\bf r}) \\
\end{matrix}
\right\}^\dag
Y^*_{LM}(\hat{\bf r}) Y_{l_\nu m_\nu}({\hat{\bf r}})
\end{eqnarray}
and can be reduced to
\begin{eqnarray} 
{\cal M}^{LM}_{l_\nu m_\nu}(k,p_{\nu_e})&=& \int d{\bf \hat r}    
\;\;Y^*_{LM}(\hat{\bf r}) Y_{l_\nu m_\nu}({\hat{\bf r}})
 \left\{
\begin{matrix}
 G_{\kappa_e }^{L,l_\nu}(k,p_{\nu_e}) \chi_{\kappa_e}^{\mu_e}(\hat{\bf r})  \\
 \\
 i F_{\kappa_e}^{L,l_\nu}(k,p_{\nu_e})\chi_{-\kappa_e}^{\mu_e}(\hat{\bf r}) \\
\end{matrix}
\right\}^\dag
\label{FG-integrals}
\end{eqnarray}
The integrals over the solid angle $d{\bf \hat r}$ are performed analytically using the orthogonal properties of the spherical harmonics. The typical integral involved in this calculation has the form:
\begin{eqnarray}
\Lambda
\left(
\begin{matrix}
\kappa_e & L & l_\nu\\
\mu_e & M & m_\nu\\
\end{matrix}
\right) = \int d{\bf \hat r} Y^*_{LM}(\hat{\bf r}) Y_{l_\nu m_\nu}({\hat{\bf r}})
\chi_{\kappa_e}^{\mu_e \dag}(\hat{\bf r}) \nonumber
\end{eqnarray} 
and explicitly reads: 
 \begin{eqnarray}
 \Lambda
 \left(
 \begin{matrix}
 \kappa_e & L & l_\nu\\
 \mu_e & M & m_\nu\\
 \end{matrix}
 \right)&=& \sqrt{ \dfrac{(2L+1)(2l_e+1)(2l_\nu +1)}{4\pi}} 
  \left(
 \begin{matrix}
 l_e & L & l_\nu\\
 0 & 0 & 0\\
 \end{matrix}
 \right) \nonumber\\ 
 &\times& \sum_{m_s} 
 C^{j_e \mu_e}_{l_e\mu_e-m_s;\frac{1}{2}m_s} \chi^{m_s \dag }  
 (-)^{M+\mu_e-m_s}   
 \left(
 \begin{matrix}
 l_e & L & l_\nu\\
 m_s-\mu_e & -M & m_\nu\\
 \end{matrix}
 \right)
\nonumber
 \end{eqnarray} 

where 
$
\left( \begin{matrix}  l_1 & l_2 & l_3\\ m_1 & m_2 & m_3\\ \end{matrix} \right) \nonumber
$
represents the $3j$ symbol with $|m_i|\le l_i$ ($i=1,2$ or $3$). This symbol is zero except for $m_1+m_2+m_3 = 0 $, $l_1+l_2+l_3 $ is integer and fulfills the triangle relation $|l_1-l_2|\le l_3 \le l_1+l_2$. These conditions altogether represent all selection rules and the law of the angular and kinetic momentum conservations.
Further, in equation (\ref{FG-integrals}) $G$ et $F$ are real functions and are given by:
\begin{widetext}
\begin{eqnarray}
\left\{
\begin{matrix}
 G_{\kappa_e}^{L,l_\nu}(k,p_{\nu_e}) \\
\\
 F_{\kappa_e}^{L,l_\nu}(k,p_{\nu_e})  \\
\end{matrix} \right\} 
= \int_0^\infty \;dr \  r^2 j_L(kr)    
j_{l_{\nu}}(p_{\nu_e} r ) \left\{
\begin{matrix}
g_{\kappa_e}(r) \\
\\
f_{\kappa_e}(r) \\
\end{matrix} \right \}.  
\label{L-integral}
\end{eqnarray}
\end{widetext}

$\beta$ spectra are usually calculated by using the electron radial wave functions evaluated at the nuclear radius $R$ \cite{harston}. This approach is not appropriate regarding the influence of the atomic effects namely the screening. For instance, the screening on the $\beta$ electron wave functions is very weak at this distance ($\phi_s(r\le R)\simeq 1 $, see equation (\ref{screening-pot}) in the next subsection below) and the corresponding modifications are completely negligible over the entire range of the spectrum. Hence, to take the screening into account, it is necessary to consider the spatial extension of the wave functions beyond the size of the nucleus by using a screened potential.\\
Consequently, the integrals (\ref{L-integral}) are performed numerically from $0$ to $\sim 5\times 10^2 R$. This limitation is due to the fact that in this study, the nucleons are described as free particles (since we are not interested in absolute decay rate) instead of bound nucleons and thus, the nucleon wave functions must vanish outside the nucleus.\\ 
\subsubsection{Screening potential}

When calculating a $\beta$ spectrum, the screening effect is generally taken into account by using a constant Thomas-Fermi potential which is subtracted from the total energy of the particle \cite{huber}. This method creates a non-physical discontinuity at the minimum energy defined by the potential \cite{lsc}. As shown in \cite{mougeot}, this simple method, which makes possible the use of analytical wave functions, is not reasonable for high $Z$ and at low energy. The usual way is to correct the Coulomb potential by the average influence of the atomic electrons. Thus, the screened potential, according to the superposition law, is given by:
\begin{eqnarray}
V(r) = -\dfrac{1}{r} -\dfrac{Z}{r} \phi_s(r)
\label{pot}
\end{eqnarray}
where $\phi_s(r) $ is the screening function which contains the average effect of the atomic electrons and the remaining $Z$ protons of the nucleus. Almost all approximate analytical expressions are derived from the Thomas-Fermi statistical model of the atom and only a few exceptions are based on self-consistent Hartree-Fock or Hartree-Fock-Slater (see \cite{GSZ}, \cite{salvat-pot} and references therein). In this paper we have used the more recent screened potentials given by Salvat equation (11) in \cite{salvat-pot} for ($Z\le 92$), i.e. 
\begin{eqnarray}
\phi_s(r) = \sum_{i=1}^3 A_i {\text e}^{-\beta_i r}
\label{screening-pot}
\end{eqnarray}
where the various parameters $A_i$ and $\beta_i$ used for the analytical screening function $\phi_s(r) $ are given by Salvat such that $\phi_s(0) = 1 $) (see Table I in \cite{salvat-pot}). For larger $Z$ we used the Moli\`ere expression given by equation (8) in the same reference \cite{salvat-pot}. Moreover, when we calculate the bound state orbitals (for larger $Z > 92 $ {\it e.g.} $^{241}$Pu) involved in the exchange corrections (see below) we have used the expression given by Green {\textit et al.} \cite{GSZ} (see also expressions (9) and (10) in \cite{salvat-pot}). Note that in the above expression (\ref{pot}) we use $(Z+1)$ as the charge of the daughter nucleus, in such a way that the asymptotic net charge, $ \lim\limits_{r\rightarrow \infty} rV(r)\rightarrow -1$, holds. Consistently, a Coulomb potential generated by a daughter nuclear charge screened by the atomic electrons was defined. 
\subsubsection{Exchange corrections}   

The calculation of the exchange effect for $^{241}$Pu has already been discussed in detail in \cite{mougeot}. It is based on the formalism given in \cite{harston}, where the exchange effect is expressed just as a correction factor which modifies the beta emission probability at low energy. Furthermore, in \cite{harston} the authors have only considered the exchange involving 
$s$ and $\bar{p}$ electrons, i.e. with $j=\frac{1}{2}$, and used for the $\beta$ electron wave function the Dirac wave function evaluated at the nuclear radius $R$ (see equation (2) of \cite{harston}). In this section, we will consider the exchange effect, which is purely a quantum effect, in a more transparent and consistent way. 

In $\beta$-electron decay measurements of a many-electron atom, it is impossible to distinguish whether the detected electron is emitted from the nucleus (the direct $\beta$-electron) or emitted from a bound state of the parent atom, while the $\beta$-electron being created in an empty bound orbital of the daughter atom. The latter case is a consequence of the the linear combination of the {\it normal product} of the two field operators associated to two electrons being created in bound and continuum final states simultaneously. The anti-commutation (fermions) relation of these fields leads to an antisymmetrization of the bound-continuum wave functions and, therefore, to additional "indirect" terms in the transition matrix element.

\noindent In the following analysis we consider a decaying parent neutral atom with $Z$ electrons whose the total wave function describing $Z$ electrons is denoted by $\Psi_{Z}({\bf r}_1,\dots,{\bf r}_Z)$. After decay, the $(Z+1)$ electrons, including the $\beta$-electron, are described by $\Psi_{Z+1}({\bf r },{\bf r}_1,\dots,{\bf r}_Z)$. Accordingly, the effects related to a decay of a many-electron atom can be included by rewriting expression (\ref{L-vect}) in a new form:
\begin{eqnarray}
L'^\mu({\bf k})= 
\int d{\bf r} \prod_{j=1}^Z d{\bf r}_j \text{e}^{i {\bf k \cdot r} } \Psi^\dagger_{Z+1}({\bf r},{\bf r}_1,\dots,{\bf r}_Z) 
\gamma^0\gamma^\mu (1-\gamma_5) \varphi_{\bar\nu_e}({\bf r}) \Psi_{Z}({\bf r}_1,{\bf r}_2,\dots,{\bf r}_Z)
\label{L-exchange1}
\end{eqnarray}
 
\noindent The expression (\ref{L-exchange1}) can be rewritten in a simpler form

\begin{eqnarray}
L'^\mu({\bf k}) = A  L^\mu({\bf k}) -
 \sum_{i \in b} B_{ai} L_i^\mu({\bf k})
\label{L-exchange2} 
\end{eqnarray}

\noindent In the above expression, the notation $ { i \in b } $ under the sum denotes all possible bound states of the daughter atom. The first term on the RHS of the equation 
(\ref{L-exchange2} ) is called the direct term, it involves the direct creation of $\beta$ electron in the continuum $\varphi_e$, while the the atomic electrons are treated as spectators during the decay process. However, the last terms are the exchange terms and related to the probability amplitude of creating an electron in a given bound subshell of the daughter atom and simultaneously accompanied by an emission of an atomic electron from an initial bound state to the continuum. The index $a$ is used to define the quantum numbers of electrons being created bound or positive energy continuum orbitals of the daughter atom.
\noindent In terms of single electron wave functions, the reduced matrix element $L_i^\mu$ in expression (\ref{L-exchange2}) can be written in the form 
\begin{eqnarray}
L_i^\mu({\bf k}) =  \int d{\bf r} \text{e}^{i {\bf k \cdot r} } 
  \bar{\varphi}_i({\bf r})\gamma^\mu (1-\gamma_5) \varphi_{\bar\nu_e}({\bf r})
\end{eqnarray}
Here $\varphi_i$ stands for the wave function of $\beta$ electron being created in a given bound subshell ($i$) of the daughter atom. Furthermore, the quantities $A$ and $B's$ are accordingly expressed as
\[
A=C \int \prod_{j=1}^Z d{\bf r}_j \left|  \varphi^\dagger_{b}({\bf r}_k) \right| 
 \left| \phi_{b}^{ }({\bf r}_k)\right|
\]
and
\[
B_{ai}= C \int \prod_{j=1}^Z d{\bf r}_j \left|  \varphi^\dagger_{a\ne i}({\bf r}_k) \right| 
\left| \phi_{b}({\bf r}_k)\right|
\]

\noindent $C$ is a normalization constant factor for a particular atom and independent of $\beta$ electron energy. $|\phi_b({\bf r}_k)|$ and $|\varphi_b({\bf r}_k)| $ stand for the Slater determinants of the single electronic wave functions describing an electron in the the orbital $b$ of the parent and daughter atom, respectively. The coefficient $A$ involves only the overlap between the electron radial wave functions of the bounds orbitals of the parent and daughter atoms. However, the exchange term, related to the coefficients $B_{ai}$ involves the overlap between the electron radial wave functions of the bound and continuum orbitals having the same quantum numbers. In addition the main contribution in the above overlaps comes from the the leading diagonal  product of the determinant \cite{harston}. Therefore, the overlaps $A$ and $B$ are related, to a good approximation, in the following expression
\[
B_{ai}= A \frac {\int  d{\bf r}' \varphi^\dagger_{e}({\bf r}') \phi_{i}({\bf r}') } 
{\int  d{\bf r}' \varphi^\dagger_{i}({\bf r}') \phi_{i}({\bf r}') }
= A \frac {\left\langle \varphi_e|\phi_i \right\rangle }{\left\langle \varphi_i| \phi_i \right\rangle}
\]
\noindent note that in the above expression, the continuum $\varphi_e$ has the same quantum numbers as the orbital $\phi_{i}$ of the parent atom from which the electron was emitted. Inserting this expression in $L'_\mu$ equation (\ref{L-exchange2}), one gets

\begin{eqnarray}
L'^\mu({\bf k}) = A \left[  L^\mu({\bf k}) -
 \sum_{i \in b}   
 \frac {\left\langle \varphi_e|\phi_i \right\rangle }{\left\langle \varphi_i| \phi_i \right\rangle}
 L_i^\mu({\bf k})\right].  
\label{L-exchange3} 
\end{eqnarray}
\noindent This equation is similar to the term in the squared bracket (equation (2) of \cite{harston}) except in this paper we have replaced $U_{E,\kappa_e}(R)$ and $U_A(R)$ wave functions evaluated at the size of the nucleus \cite{harston} with respectively $L(k)$ and $L_i(k)$ containing the integration over the entire space-position of the $\beta$ electron. Secondary, in \cite{harston} the exchange terms contain only those terms with $j = 1/2$, while in our consideration we include all possible exchanges in the expression.\\  
\noindent $\phi_i$ are the single electron bound wave functions, solutions of the  Dirac equation which describes a bound electron in the field of the parent nucleus with a charge $Z$ and the field of the remaining ($Z-1$) bound electrons. In contrast to the potential (\ref{pot}) used to obtain the continuum wave functions ($E\ge mc^2$), the potential here accounts for the electrostatic self-interaction which must be subtracted from (\ref{pot}) (see \cite{salvat-pot} for more details).
In this paper, we use the analytical potentials given by Salvat (see equations (4), (6), (11) and (12) of \cite{salvat-pot}) for ($Z\le 92$). And, for larger $Z$ we use the expressions (9) and (10) from the same reference \cite{salvat-pot} and the two parameters in the cited expressions are obtained from \cite{GSZ}.

Here $\varphi_e$ and $\varphi_b$ stand for a single electron continuum and bound Dirac wave functions in the field of the daughter atom, respectively. As stated above, in order to obtain the bound wave function $\varphi_b$ describing a single electron in an atomic orbital, one has to account for self-interaction and to remove it from the screened potential of the daughter nucleus with a charge $Z+1$.

\section{Results and discussion}
\label{secres}

In this section we will briefly consider the beta spectra of $^{241}$Pu \cite{Pu-Expr}, $^{63}$Ni \cite{Ni-Expr} and $^{138}$La \cite{La-Expr}. This consideration is based on the results derived in the previous sections where 
the atomic exchange and the Coulomb effects due to the nuclear point-like charge screened by the atomic electrons were included. The numerical calculations are then compared to recent experimental data.
The $\beta$ spectra of $^{241}$Pu and $^{63}$Ni were recently measured using metallic magnetic calorimeters. Such detectors have been shown to be a powerful experimental technique to study the energy spectra of beta particles having a maximum energy of about 1 MeV. In $^{63}$Ni $\beta$-decay, the transition is allowed ($\frac{1}{2}^- \rightarrow  \frac{3}{2}^-$), while for $^{241}$Pu the transition is first forbidden non unique  ($\frac{5}{2}^+ \rightarrow  \frac{5}{2}^-$). In the case of $^{241}$Pu, although the transition is first forbidden non unique, it can be calculated as allowed in the framework of the $\xi$ approximation \cite{mougeot}. Moreover, complete calculation of the nuclear matrix elements in the framework of the Nilsson model was already performed in \cite{rizek}. It was shown, that the corresponding correction increases linearly with the energy, not higher than 0.3$\%$ at the endpoint and 0.1$\%$ below 8 keV. 
Thus, this transition can be calculated with confidence as allowed. The $^{138}$La decay exhibits a second forbidden unique transition (${5}^+ \rightarrow {2}^+$) and has been measured in a very recent experiment using a LaBr$_3$:Ce scintillator \cite{La-Expr}, where enhanced counting statistics and a low energy cut off of 2.5~keV were achieved. All these measured spectra, with high accuracy, might be an excellent challenge for the theoretical calculations. In the present work, these experiments are compared with our calculations with and without screening and exchange corrections.

\begin {figure}
\begin{center} 
\includegraphics[scale=1.1]{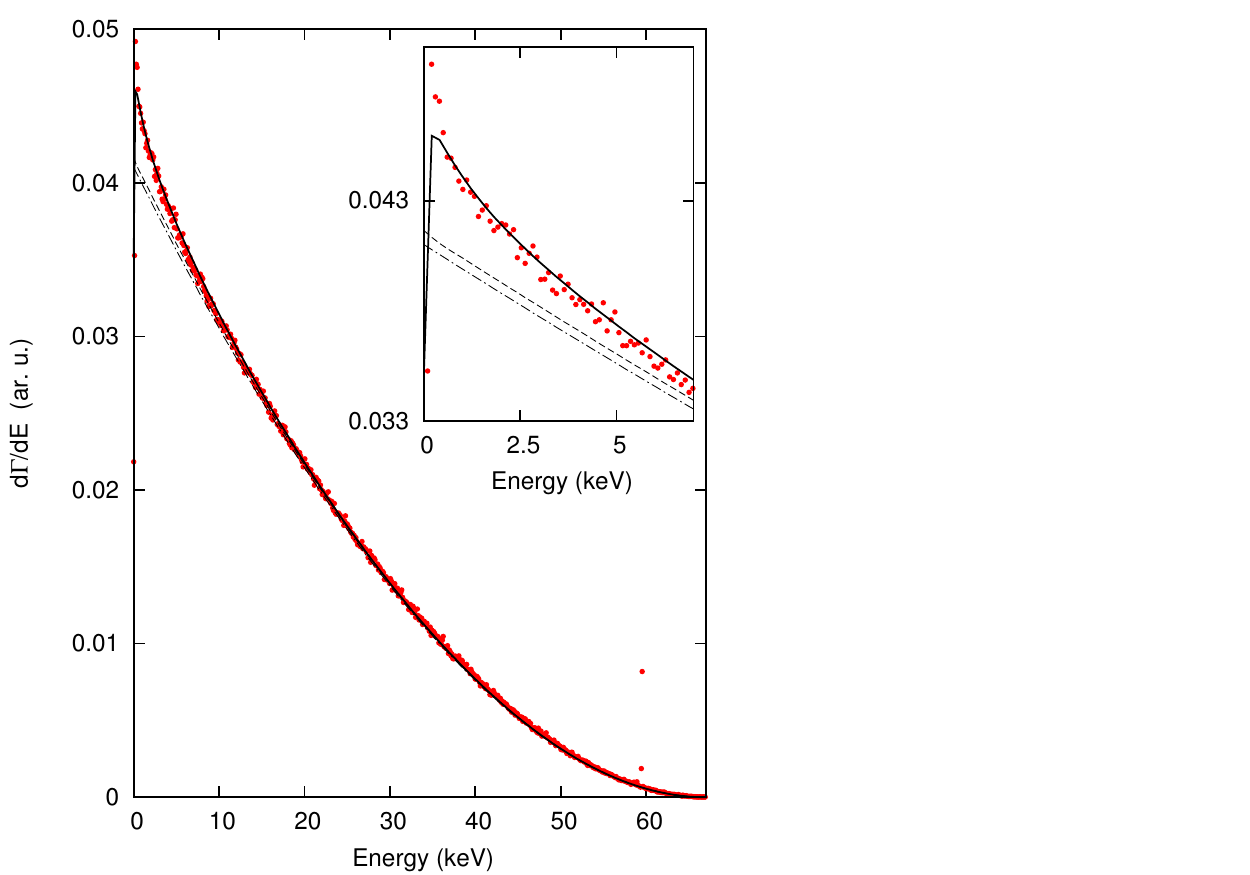}
\end{center}
\caption{Energy spectrum of emitted $\beta$-electron in $^{63}$Ni decay. Solid circles show measured data \cite{Ni-Expr}. Solid line presents calculation including the screening and the exchange corrections. Dashed line represents the screening contribution only. Dot-dashed line is the result of using only the simple Coulomb correction potential wave functions without screening. The panel inset highlights different calculations at low energy. The intensity of the spectrum is given in arbitrary units (ar. u.), and therefore, the theoretical calculations are normalized to the experimental data at high energies.}
\label{fig1}
\end {figure}

In FIG.\ref{fig1} we illustrate a comparison between calculated spectra and the experimental data of $^{63}$Ni. The theoretical calculations are normalized to the experimental data at high energies. This normalization is due to the fact that we have used the Dirac plane waves to describe the nucleons inside the nucleus. The figure shows that the screening in comparison with pure Coulomb description for $^{63}$Ni has a very tiny effect, as expected for this low $Z$ nucleus. However the exchange corrections have a valuable influence on the energy spectra at low emission energy. One can see a very good agreement when both, screening and exchange, corrections are taken into account. The low contribution of the screening correction is due to the fact that the size of the nucleus is much smaller than the size of the inner atomic orbital ($\sim Z^{-1}$), thus much smaller than the size of the atom itself. Therefore, when spatial integration involved in the lepton current is performed within a region much smaller than the size of the atom, the screened potential is practically reduced to the pure Coulomb field: $\phi_s (r\ll Z^{-1})\simeq 1$. On the other hand, when calculating the exchange contribution to the spectrum, the atomic orbitals are calculated in a screened potential, which must distort the wave functions and can contribute significantly to the overlap between the orbitals of the parent and daughter atoms. This does not concern the wave functions of the $\beta-$electron evaluated within a small region of order of $10^2 R$.
\begin {figure}
\begin{center} 
\includegraphics[scale=1.1]{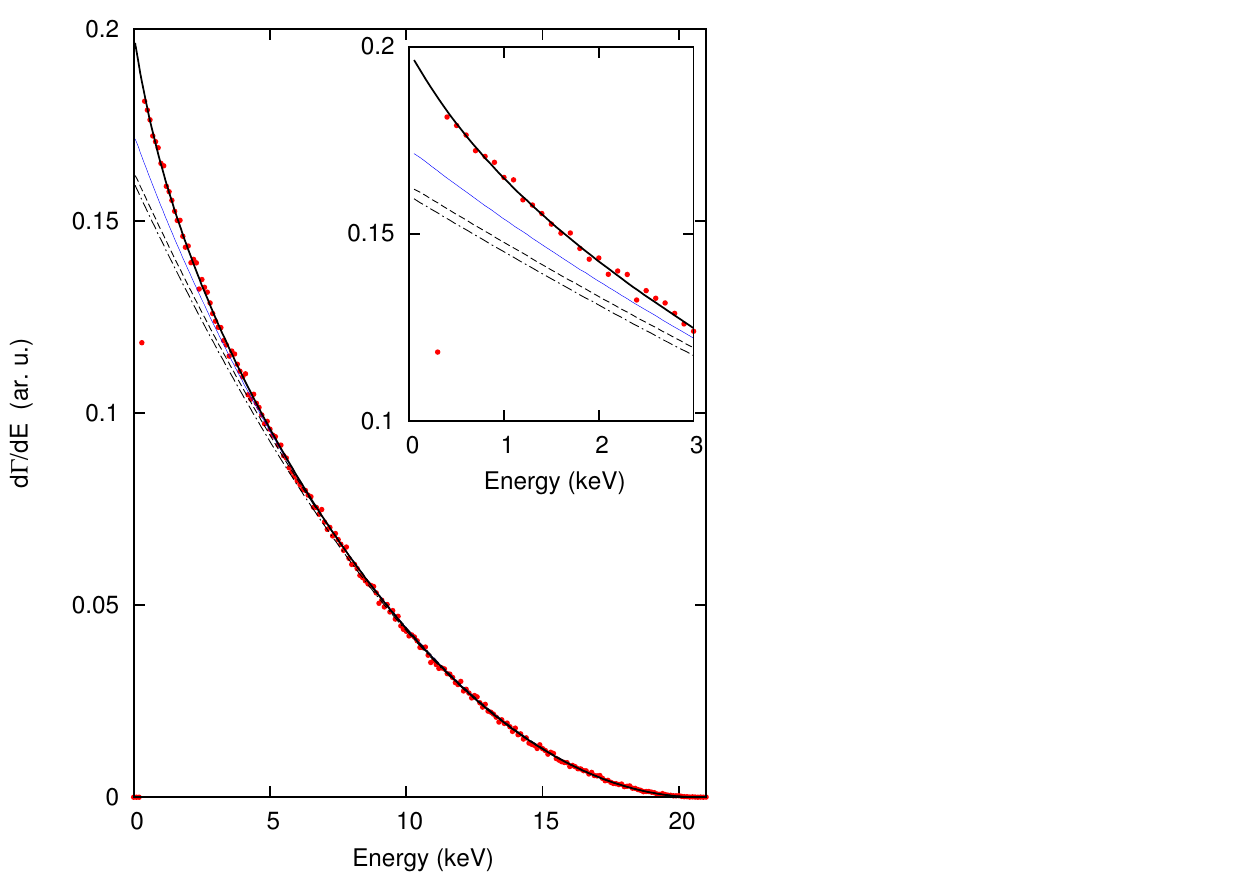}
\end{center}
\caption{The same as FIG.\ref{fig1} but for $^{241}$Pu. The experimental data are from \cite{Pu-Expr}. The blue-solid curve corresponds to partial exchange, where only the exchange with states $n\le 2$. Theoretical calculations are normalized to the experimental data at high energies.}
\label{fig2}
\end {figure}

However, in the case of heavier radionuclide such as $^{241}$Pu, one could expect an important screening effect since the size of the inner atomic orbitals is smaller ($\sim Z^{-1}$) and the electronic cloud close to the nucleus. In FIG.\ref{fig2} we illustrate the comparisons between calculated spectra and the experimental data as in FIG.\ref{fig1} but for $^{241}$Pu. Theoretical calculations are normalized to the experimental data as in FIG.\ref{fig1}. The figure shows that the effects due to the screening in the wave function compared to the Coulomb description are less pronounced, as in the case of $^{63}$Ni, in contrast to the explanation given above. However, the exchange correction has a substantial influence on the energy spectrum and the agreement with the experimental data is more reproduced in the low energy region. Although, the screening, on one's own, seems to have a tiny effect on the spectra, its combination with the exchange correction contributes very significantly to the $\beta$ spectra through the distortion of the atomic orbitals in the screened potential, and thus, modifies significantly the overlap between the atomic orbitals before and after the decay. Furthermore, calculated spectra shown in FIG.\ref{fig1} and FIG.\ref{fig2} were obtained by performing the radial integration over a region from 0 to $R_{max}$ and the results therefore obtained are $R_{max}$ independent, provided that $R_{max}\le 5\times10^2 R$.  

\begin {figure}
\begin{center} 
\includegraphics[scale=1.1]{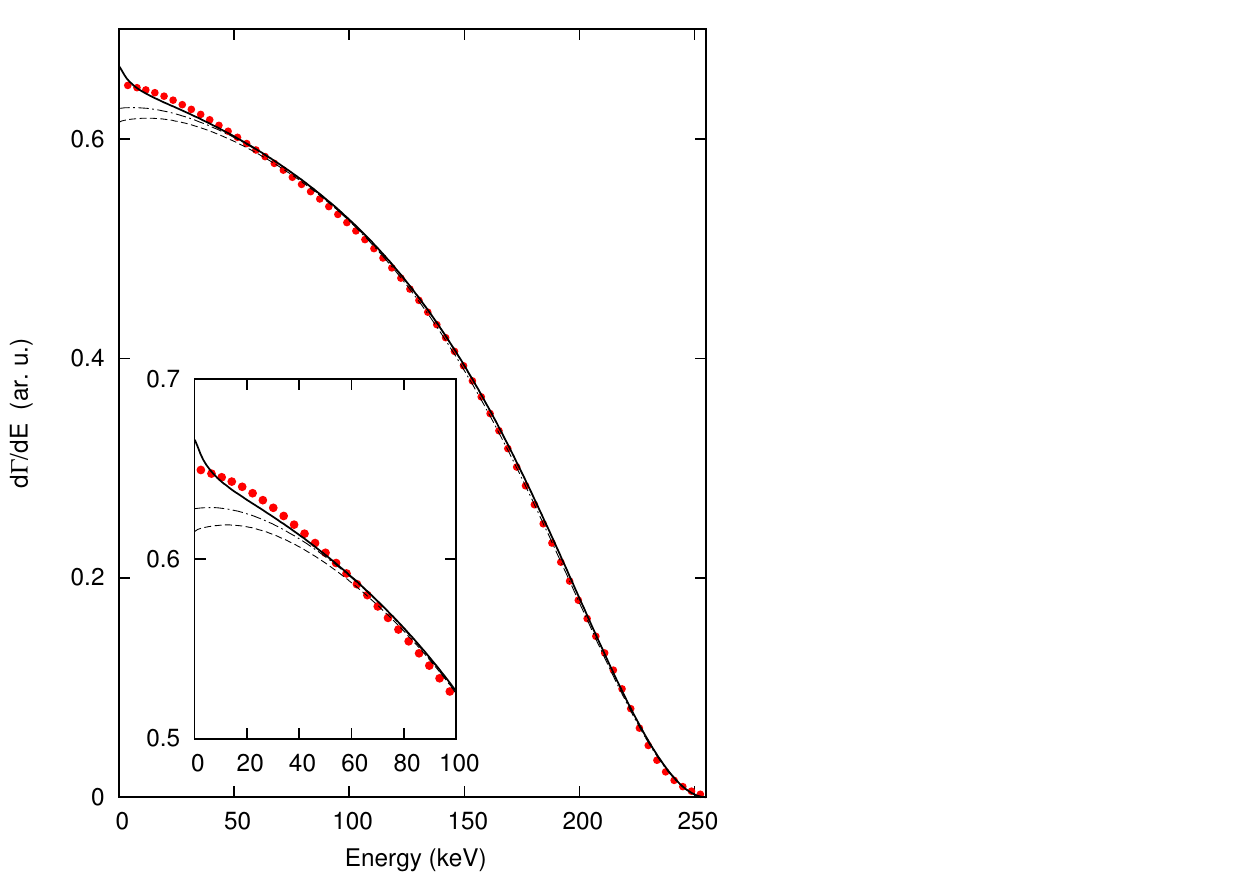}
\end{center}
\caption{The same as FIG.\ref{fig1} but for $^{138}$La. The experimental data are from \cite{La-Expr}. The theoretical calculations are normalized to the experimental data at high energies.}
\label{fig3}
\end {figure}

FIG.\ref{fig3} illustrates our theoretical calculations for the second forbidden unique transition of $^{138}$La ($L=2$, $\Delta J = L+1$ and $\pi_i . \pi_f = (-1)^L $) compared to recent experimental data \cite{La-Expr}. As for both allowed and first non unique forbidden transitions, the figure shows that the effects due to the screening in the $\beta-$electron wave function are negligible. Furthermore, the exchange and screening effects together have a very significant influence on the energy spectrum. Nevertheless some discrepancy remains at low energies (below 5 keV) where theoretical calculations with screening and exchange predict a strong decrease in the spectrum in contrast to measured data. In addition, we met a serious complication in the calculation of the spectrum of $^{138}$La due to the sensitivity of the results on the size of the region of integration. The above results displayed in FIG.\ref{fig3} were obtained for $4.4\times10^2 R \le R_{max}\le 4.5\times10^2 R$.   

\section{Conclusion}

In conclusion, we have considered the energy spectra of emitted electron in $\beta$-decay. We have focused our attention on the exact determination of the leptonic vector current. The corrections due to the screened Coulomb field by the atomic electron cloud were included. It is shown that, as a result of these corrections, the low energy region of the $\beta$ spectrum is slightly modified. The indirect term, or so-called exchange contribution, has been included consistently to the transition matrix elements of the $\beta$-decay. The latter correction can be thought of as a consequence of the antisymmetrization of the bound atomic electrons and the continuum $\beta$-electron wave functions.\\ 
In such a case, we have shown that the exchange effects have a strong influence on the energy spectra at low energy region. Comparisons between calculated spectra and recent experimental data of $^{63}$Ni, $^{241}$Pu and $^{138}$La decays show very good agreements when both screening and exchange corrections are included. The agreement is excellent down to very low emission energies for both $^{63}$Ni and $^{241}$Pu, independently on the region of integration, provided that $R_{max}\le 5\times 10^2 R$.\\ 
A small discrepancy remains in the spectrum of the second forbidden unique transition of $^{138}$La. It can be due to the quality of the measurement from \cite{La-Expr}. Quarati \textit{et al.} are currently carrying out new measurements of the $^{138}$La decay, and there are some clues for a rise of the $\beta$ spectrum below 10~keV. This discrepancy can also be thought of as occurring due to the crude description of the nuclear current. Indeed, in this paper we have assumed that the active nucleons, during the decay process, move freely in the nucleus. Thus a different and more elaborate description of the nuclear current is required and will be considered in a future paper. 

\section*{ACKNOWLEDGEMENTS}

The authors would like to thank A. B. Voitkiv, J. P. Engel J. Dudek and H. Molique for numerous and helpful discussions and M. Loidl, C. Le Bret  and F.G.A. Quarati for providing their experimental work.

 

\begin{thebibliography}{99}
\bibitem{gammas} In The convention used in this paper the $\gamma$'s are connected to the usual $\beta$ and $\alpha$'s Dirac matrices by the relations $\gamma^0 = \beta $ and $\gamma^i = \beta \alpha_i$ ($i=1,2$ or $3$ ). Further, $\gamma^5 = i\gamma^0\gamma^1\gamma^2\gamma^3 $.
\end{thebibliography}
\end{document}